\RequirePackage{ifpdf}
\ifpdf 
\documentclass[pdftex]{sigma}
\else
\documentclass{sigma}
\fi

\begin{document}

\allowdisplaybreaks

\renewcommand{\PaperNumber}{024}

\FirstPageHeading

\renewcommand{\thefootnote}{$\star$}

\ShortArticleName{Quark Interactions on Dynamical Chiral Symmetry Breaking by a Magnetic Field}

\ArticleName{Ef\/fects of Quark Interactions on Dynamical Chiral\\ Symmetry Breaking by a Magnetic Field\footnote{This paper is a contribution to the
Proceedings of the Seventh International Conference ``Symmetry in
Nonlinear Mathematical Physics'' (June 24--30, 2007, Kyiv,
Ukraine). The full collection is available at
\href{http://www.emis.de/journals/SIGMA/symmetry2007.html}{http://www.emis.de/journals/SIGMA/symmetry2007.html}}}

\Author{Brigitte HILLER~$^\dag$, Alexander A. OSIPOV~$^\ddag$, Alex H. BLIN~$^\dag$ and Jo\~ao da PROVID\^ENCIA~$^\dag$}

\AuthorNameForHeading{B. Hiller, A.A. Osipov, A.H. Blin and J. da Provid\^encia}

\Address{$^\dag$~Centro de F\'{\i}sica Te\'{o}rica, Departamento de F\'{\i}sica da Universidade de Coimbra,\\
 $\phantom{^\dag}$~3004-516 Coimbra, Portugal}
\EmailD{\href{mailto:brigitte@teor.fis.uc.pt}{brigitte@teor.fis.uc.pt}, \href{mailto:alex@teor.fis.uc.pt}{alex@teor.fis.uc.pt}, \href{mailto:providencia@teor.fis.uc.pt}{providencia@teor.fis.uc.pt}}

\Address{$^\ddag$~Joint Institute for Nuclear Research, Laboratory of Nuclear Problems,\\
 $\phantom{^\ddag}$~141980 Dubna, Moscow region, Russia}
\EmailD{\href{mailto:osipov@nu.jinr.ru}{osipov@nu.jinr.ru}}

\ArticleDates{Received November 14, 2007, in f\/inal form February
07, 2008; Published online February 22, 2008}

\Abstract{It is shown how the strong interaction dynamics of a multi-quark Lagrangian af\/fects the catalysis of dynamical symmetry breaking by a constant magnetic f\/ield in $(3+1)$ dimensions. Attention is drawn to the local minima structure of the theory.}


\Classification{81T10}

\renewcommand{\thefootnote}{\arabic{footnote}}
\setcounter{footnote}{0}

\section{Introduction}

The existence of a zero-energy surface in the spectrum of a Dirac
particle is ensured for any homogeneous magnetic f\/ield with a
f\/ixed direction by a quantum mechanical supersymmetry of the
corresponding second-order Dirac Hamiltonian \cite{Jackiw:1984}.
For interacting fermions subject to an attractive two-body force in $2+1$ and $3+1$ dimensions, a constant magnetic
f\/ield catalyzes the dynamical symmetry breaking leading to fermion pairing and to a
fermion mass even for arbitrarily weak coupling strength
\cite{Lemmer:1989,Klimenko:1991,Krive:1991}.
The zero-energy surface of the lowest
Landau level (LLL) plays a crucial role in the dynamics of such fermion
pairing \cite{Gusynin:1994,Miransky:1995,Miransky:1996}, and the
generated fermion mass, $M_{\rm dyn}$, turns
out to be much smaller than the Landau gap $\sim\sqrt{|eH|}$.  The dynamics
of the fermion pairing in the homogeneous magnetic f\/ield is
essentially $(1+1)$-dimensional, and the deep analogy of this phenomenon
with the dynamics of electron pairing in BCS \cite{Bardeen:1957} has
been stressed.
The generation of mass via catalysis does however not prevail for arbitrary  magnetic f\/ield conf\/igurations. For instance, it has been demonstrated in \cite{Ragazzon:1994,Ragazzon:1999} that the Nambu--Jona-Lasinio (NJL)
model \cite{Nambu:1961} minimally coupled to a background magnetic
f\/ield with variable direction becomes massive only beyond
some critical value of the strong coupling constant. In these cases dynamical chiral symmetry breaking corresponds to the standard scenario.

In the present paper we show and discuss the inf\/luence that the structure of interaction terms, specif\/ied below, may have on the ef\/fect of catalysis by a constant magnetic f\/ield (from now on we use interaction terms to denote interaction among the fermions, the interaction with the magnetic f\/ield will be explicitly referred as such). We obtain that catalysis of dynamical symmetry breaking always occurs at arbitrarily small magnetic f\/ields for a system which is subcritical in the interaction couplings in absence of the magnetic f\/ield. However for increasing strength of the magnetic f\/ield and depending on the values of interaction couplings, another minimum of the ef\/fective potential emerges at larger dynamical fermion masses, which competes with the catalyzed minimum and eventually becomes the globally stable state of the system at characteristic scales $H=10^{19}$ Gauss. The broken symmetry is not restored with increasing strength of the magnetic f\/ield.

A few words about the fermion interaction model used in this paper: in recent years we have reported on the importance of adding eight-quark interaction terms \cite{Osipov:2005} to the  four-quark $U(3)_L\times U(3)_R$ chiral symmetric NJL Lagrangian joined with the $U(1)_A$ breaking 't Hooft six-quark interaction \cite{Hooft:1978}. Their inclusion leads to a scalar ef\/fective potential which is globally stable and to a full equivalence between the stationary phase approach (SPA) and the mean f\/ield (MF) approximation methods used to obtain it. Without them the model is af\/f\/licted by an unstable (SPA) respectively metastable (MF) vacuum
\cite{Osipov:2006a}, as presented two years ago at the Kyiv 2005 conference \cite{Osipov:2006b}. There we have also shown that going beyond the leading order results of SPA did not cure the instability of the ef\/fective potential. Although we prefer to consider the inclusion of eight quark interactions as the minimal chiral approach to render the vacuum of the NJL model with 't Hooft interaction stable, QCD regarded as an ef\/fective theory displays an inf\/inite chain of multiquark interactions
\cite{Zwanziger:2006}. The issue of the corresponding $N_c$  counting rules is however still open.  In the framework of the instanton gas model beyond the zero mode approximation multiquark interactions are delivered at any $2n$-quark order $(n\ge 2)$  with equal weight in the large $N_c$ limit \cite{Simonov:2002}. On the other hand lattice calculations of gluon correlators \cite{Bali:2001} give a larger  weight to the lower correlators. This leads to the expectation that after integrating out the gluonic degrees of freedom, a similar hierarchy could arise for the multiquark interactions. In our approach higher order interactions are assumed to be suppressed in the large $N_c$ counting~\cite{Osipov:2005,Osipov:2007}.

Since then the eight-quark extended model has been carefully studied in connection with the low lying spectra and characteristics of the pseudoscalar and scalar nonets \cite{Osipov:2007}, temperature dependence of chiral transitions \cite{Osipov:2007a}, and \cite{Sakaguchi:2007} for the two f\/lavor case, and coupling to a constant magnetic f\/ield \cite{Osipov:2007b}. We have learned that the ef\/fects of the eight quark interactions on the mass spectra are not essential, except for the mass of the lowest state of the mixed singlet-octet scalar states, which decreases with increasing strength of the eight quark interaction terms that violate the Okubo--Zweig--Iizuka (OZI) rule \cite{Okubo:1963}. However, their ef\/fect can be dramatic on studies involving changes of the extrema of the ef\/fective potential, which are for instance the cases related with the thermodynamics, the in medium dependence, and the coupling to electromagnetic f\/ields of the multi-fermion system. The latter is the main subject of the present contribution. This work is very much based on our recent paper~\cite{Osipov:2007b}, it is however a more extensive version. It contains also truly novel calculations, as we chose as starting conf\/igurations a regime of completely sub-critical solutions (i.e.\ only the trivial vacuum with vanishing condensate exists in absence of the magnetic f\/ield), which were not analyzed before for the enlarged multi-quark Lagrangian. We study then how the system responds to coupling to a magnetic f\/ield, starting with an arbitrarily weak strength and increasing it successively.
First appear the well known catalyzed solutions as minima of the gap equation. Then the fate of the system may be either a f\/irst order transition or a crossover to heavier condensates, depending on the strength of the couplings of multiquark interactions.
The paper is structured as follows. In Section~\ref{sec2} we present the model Lagrangian in presence of a constant magnetic f\/ield and the corresponding ef\/fective potential. In Section~\ref{sec3} we obtain the solutions of the gap equations and their relation with the Landau levels and highlight the dif\/ferences to the well-known cases. In Section~\ref{sec4} we present the conclusions.

\newpage

\section{Model Lagrangian}\label{sec2}

In the hadronic QCD vacuum the relevant fermionic interactions can be
modeled by the well-known Nambu--Jona-Lasinio (NJL) Lagrangian
\cite{Nambu:1961}. The coupling to a constant magnetic f\/ield has been done
in \cite{Miransky:1995} for  one-f\/lavour fermions.

For a more realistic analysis one should take into account the
dif\/ferent quark f\/lavours of the QCD vacuum. These
extensions of the NJL model are well-known
\cite{Eguchi:1976,Volkov:1982,Ebert:1986}, for instance, the
four-quark $U(3)_L\times U(3)_R$ chiral symmetric Lagrangian together
with the $U(1)_A$ breaking 't Hooft six-quark interactions has been
extensively studied at the mean-f\/ield level
\cite{Bernard:1988,Reinhardt:1988,Klimt:1990,Bernard:1993,Hatsuda:1994}.
As mentioned in the introduction, it has been recently shown \cite{Osipov:2005} that
the eight-quark interactions are of vital importance to stabilize the
multi-quark vacuum.

The multi-quark dynamics of the extended NJL model is described by
the Lagrangian density
\begin{gather*}
  {\cal L}_{\rm ef\/f} =\bar{q}(i\gamma^\mu D_\mu - \hat{m})q
          +{\cal L}_{4q} + {\cal L}_{6q} + {\cal L}_{8q},
\end{gather*}
where the gauge covariant derivative $D_\mu$ is equal to $D_\mu =
\partial_\mu +iQA_\mu$ with the external electromagnetic f\/ield
$A_\mu$ and quark charges $Q=e\cdot\mbox{diag}(2/3,-1/3,-1/3)$. We consider a constant
magnetic f\/ield,
$A_x=-Hy$, $A_y=A_z=0$ (Landau gauge). It is
assumed that quark f\/ields have colour $(N_c=3)$ and f\/lavour
$(N_f=3)$ indices. The current quark mass, $\hat{m}$, is a diagonal
matrix with elements $\mbox{diag} (\hat{m}_u, \hat{m}_d, \hat{m}_s)$,
which explicitly breaks the global chiral $SU_L(3)\times SU_R(3)$
symmetry of the Lagrangian. We shall neglect this ef\/fect in the
following assuming that $\hat{m}=0$.

The most general form of multi-quark interactions in the scalar and pseudoscalar channels
are of the form
\begin{gather*}
  {\cal L}_{4q}   =  \frac{G}{2}\left[(\bar{q}
  \lambda_aq)^2+ (\bar{q}i\gamma_5\lambda_aq)^2\right], \!\!\qquad
  {\cal L}_{6q}   =  \kappa (\det \bar{q}P_Lq
                      + \det \bar{q}P_Rq),\!\! \qquad
  {\cal L}_{8q}  =  {\cal L}_{8q}^{(1)} +
                           {\cal L}_{8q}^{(2)}.
\end{gather*}
The $U(3)$ f\/lavour matrices $\lambda_a$, $a=0,1,\ldots ,8,$
are normalized such that $\mbox{tr} (\lambda_a \lambda_b) =
2\delta_{ab}$. The matrices $P_{L,R}=(1\mp\gamma_5)/2$ are
chiral projectors and the determinant is over f\/lavour indices,
which are suppressed here. The determinantal interaction breaks
explicitly the axial $U(1)_A$ symmetry \cite{Hooft:1978} and Zweig's
rule. The eight-quark spin zero interactions are given by
\begin{gather*}
   {\cal L}_{8q}^{(1)}=
   8g_1\left[ (\bar q_iP_Rq_m)(\bar q_mP_Lq_i) \right]^2, \qquad
   {\cal L}_{8q}^{(2)}=
   16 g_2 (\bar q_iP_Rq_m)(\bar q_mP_Lq_j)
   (\bar q_jP_Rq_k)(\bar q_kP_Lq_i).
\end{gather*}
$G$,  $\kappa$, $g_1$, $g_2$ are dimensionful coupling constants:
$[G]=M^{-2}$, $[\kappa ]=M^{-5}$, $[g_1]=[g_2]=M^{-8}$ in units
$\hbar =c=1$. The term proportional to $g_1$ violates the OZI rule.

\subsection[Effective potential]{Ef\/fective potential}\label{sec2.1}

We proceed by calculating the ef\/fective potential of the
theory, $V(m_u, m_d, m_s)$, in the constant magnetic f\/ield
$A_x=-Hy$, $A_y=A_z=0$. The arguments, $m_i$, are
simply real parameters; they are not to be identif\/ied with the
masses of any presumed one-particle states. Instead, we shall use the
capital letter $M_i$ for the point where $V$ takes its local minimum,
which specif\/ies the masses of constituent quark f\/ields.

The potential is built of the following two terms
\begin{gather}
\label{efpot}
   V(m_u,m_d,m_s)=V_{\rm st}+V_{S}.
\end{gather}
The f\/irst contribution results from the many-fermion vertices
of Lagrangian ${\cal L}_{\rm ef\/f}$, after reducing them to a bilinear form
with help of bosonic auxiliary f\/ields, and subsequent
integration over these f\/ields, using the stationary phase method.
The specif\/ic details of these calculations and the result are
given in~\cite{Osipov:2005}. We obtain
\begin{gather*}
     V_{\rm st} = \frac{1}{16}
     \left( 4Gh_i^2  + \kappa h_uh_dh_s + \frac{3g_1}{2}
     \left(h_i^2\right)^2 +3g_2 h_i^4\right),
\end{gather*}
where $h_i^2=h_u^2+h_d^2+h_s^2$, and $h_i^4=h_u^4+h_d^4+h_s^4$. The
functions $h_i$ depend on the coupling constants $G$, $\kappa$, $g_1$, $g_2$
and on the independent variables $\Delta_i=m_i-\hat{m}_i$. To
f\/ind this dependence one should solve the system of cubic
equations
\begin{gather}
   Gh_u + \Delta_u +\displaystyle\frac{\kappa}{16}  h_dh_s
   +\displaystyle\frac{g_1}{4}  h_u h_i^2
   +\displaystyle\frac{g_2}{2}   h_u^3=0, \nonumber\\
   Gh_d + \Delta_d +\displaystyle\frac{\kappa}{16} h_uh_s
   +\displaystyle\frac{g_1}{4} h_d h_i^2
   +\displaystyle\frac{g_2}{2} h_d^3=0, \label{saddle-1}\\
   Gh_s + \Delta_s +\displaystyle\frac{\kappa}{16} h_uh_d
   +\displaystyle\frac{g_1}{4} h_s h_i^2
   +\displaystyle\frac{g_2}{2} h_s^3=0.\nonumber
   \end{gather}
In the parameter range
\begin{gather}
\label{ineq1}
   g_1>0, \qquad g_1 +3g_2>0, \qquad
   G>\frac{1}{g_1}\left(\frac{\kappa}{16}\right)^2.
\end{gather}
the system has only one set of real solutions,
and this guarantees the vacuum state of the theory to be stable~\cite{Osipov:2005}.

The second term on the r.h.s.\ of equation~(\ref{efpot}) derives from the
integration over the quark bilinears in the functional integral of the
theory in presence of a constant magnetic f\/ield $H$. As has
been calculated by Schwinger a long time ago~\cite{Schwinger:1951}
\begin{gather*}
     V_{S} = \sum_{i=u,d,s} V_S(m_i, |Q_iH|),
\end{gather*}
where
\begin{gather*}
 V_S(m, |QH|)   =\frac{N_c}{8\pi^2}\int\limits_0^\infty
   \frac{d s}{s^2} e^{-sm^2}\rho (s,\Lambda^2)
   |QH|\coth (s|QH|),
\end{gather*}
up to an unessential constant.
Here the cutof\/f $\Lambda$ has been introduced by subtracting
of\/f suitable counterterms to regularize the integral at the
lower limit, i.e., $\rho (s,\Lambda^2)=1-(1+s\Lambda^2)
e^{-s\Lambda^2}$. For the fermion tadpole this works as the
four-momentum covariant cutof\/f in the Euclidean space:
$\vec{p}\,^2+p_4^2<\Lambda^2$. As a result we obtain
\begin{gather*}
 V_S(m, |QH|)   =\frac{N_c}{8\pi^2}\left\{\Lambda^2|QH|\left[\ln 2\pi
      - 2\ln\Gamma\left(\frac{\Lambda^2+m^2}{2|QH|}\right)\right] +m^2|QH|\ln\left(1+\frac{\Lambda^2}{m^2}\right)
      \right.\nonumber \\
 \phantom{V_S(m, |QH|)   =}{} +4(QH)^2
      \frac{d}{d\nu}\left[ \zeta\left(\nu -1,
   \frac{\Lambda^2+m^2}{2|QH|}\right)-
   \left.\zeta\left(\nu -1, \frac{m^2}{2|QH|}\right)\right]
   \right|_{\nu =0}   \nonumber \\
 \left.  \phantom{V_S(m, |QH|)   =}{}+\frac{\Lambda^4}{2}\left(\ln\frac{\Lambda^2}{2|QH|}
   -\frac{3}{2}\right)-\Lambda^2m^2\right\}.
\end{gather*}
The quantity $\zeta (\nu ,x)$ denotes the generalized Riemann zeta
function \cite{Bateman:1953}.

\section{Analysis of solutions}\label{sec3}

In this section we study the extrema of the ef\/fective potential, given by the gap equations. We start by considering the known case with four-quark interactions in Subsection \ref{sec3.1}. A detailed study of several relevant limits and the role of Landau levels in the catalysis of dynamical symmetry breaking is presented. Then in Subsection~\ref{sec3.2} the extended system with six and eight quark interactions is analyzed and its new features discussed.

\subsection[The standard four-quark system in a constant magnetic field]{The standard four-quark system in a constant magnetic f\/ield}\label{sec3.1}

We consider f\/irst the simple $SU(3)$ f\/lavour
limit for the situation in which $\hat m=0$ and $\kappa =g_1=g_2=0$.
For purposes of illustration, we ignore in the remaining the charge
dif\/ference of $u$ and~$d$,~$s$ quarks. The averaged common charge
$|Q|=|4e/9|$ will be used. In this case one gets the potential
$V(m)=N_f(m^2/4G+V_S(m, |QH|))$\footnote{Note that the $SU(3)$ result for the
      ef\/fective potential divided by the number of
      f\/lavours does not coincide with
      the one f\/lavour
      case. This is due to group structure factors.} and the gap equation is obtained as
\begin{gather*}
   \frac{d V(m)}{d m}= mN_c N_f\Bigg\{ \frac{1}{2GN_c}
   -\frac{\Lambda^2}{4\pi^2} \psi\left(
   \frac{\Lambda^2+m^2}{2|QH|}\right)   \nonumber \\
\phantom{\frac{d V(m)}{d m}=}{} + \frac{|QH|}{4\pi^2} \left[J_1(m^2)+2\ln
   \frac{ \Gamma\left( \frac{\Lambda^2+m^2}{2|QH|} \right)
   }{\Gamma \left( \frac{m^2}{2|QH|} \right)}\right]\Bigg\}=0,
\end{gather*}
where $\psi (x)=d\ln\Gamma (x)/d x$ is the Euler dilogarithmic
function, and
\begin{gather*}
   J_1(m^2)=\ln\left(1+\frac{\Lambda^2}{m^2}\right)
           -\frac{\Lambda^2}{\Lambda^2+m^2}.
\end{gather*}

\begin{figure}[t]
\centerline{\includegraphics[width=6.74cm]{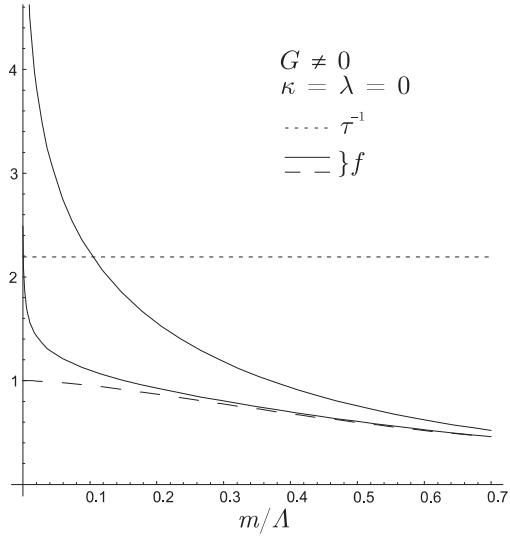}}
\vspace{-2mm}

\caption{The the l.h.s.\ (straight short-dashed line) and the r.h.s.\ (other curves) of the gap equation~(\ref{gap2}) related to the four-quark Lagrangian, as functions of $m/\Lambda$, all in dimensionless units. Three
dif\/ferent values of the magnetic f\/ield strength $H$ are considered: the long-dashed curve corresponds to $H=0$; the two full curves represent the cases for magnetic f\/ields of strength $|QH|\Lambda^{-2}=0.5$ (upper curve) and
$|QH|\Lambda^{-2}=0.1$. For any $H\ne0$ the l.h.s.\ and r.h.s.\ intersect and yield the non-trivial solution of~(\ref{gap2}). For $H=0$ the system is subcritical for the parameter set considered. It becomes critical in the case $\tau >1$, where
$\tau =G\Lambda^2N_c/2\pi^2$. }
\label{fig1}\vspace{-1mm}
\end{figure}

The gap equation always has a trivial solution, $m=0$. The
nontrivial solution is contained in the equation
\begin{gather}
   \frac{2\pi^2}{G\Lambda^2N_c} = f(m^2;\Lambda,|QH|) \equiv
   \psi\left(\frac{\Lambda^2+m^2}{2|QH|}\right)
-\frac{|QH|}{\Lambda^2}\left[J_1(m^2)+2\ln
   \frac{ \Gamma\left( \frac{\Lambda^2+m^2}{2|QH|} \right)
   }{\Gamma \left( \frac{m^2}{2|QH|} \right)}\right].\label{gap2}
\end{gather}
Note that although our regularization scheme
dif\/fers essentially from the one used in \cite{Miransky:1995},
the main conclusion of that paper is not changed: as in
\cite{Miransky:1995}, our gap equation has a nontrivial solution at
all $G>0$, if $H\neq 0$. Fig.~\ref{fig1} illustrates this important result.
Displayed are the r.h.s.\ (curves~$f$) and l.h.s.\ of equation~(\ref{gap2}), shown only
for the positive valued abscissa, since the curves are symmetric. The
numerical value of the four-quark coupling is taken to be
$G \Lambda^2=3$. The l.h.s.\ is a constant and indicated by the
short-dashed line. The r.h.s.\ is shown as the long-dashed curve
(for $H=0$) and as full lines for $|QH|\Lambda^{-2}=0.1$ and
$|QH|\Lambda^{-2}=0.5$. One sees that in absence of the magnetic
f\/ield one is in the subcritical regime of dynamical symmetry
breaking, for this choice of $G \Lambda^2$. The inclusion of a
constant magnetic f\/ield, however small it might be, changes radically
the dynamical symmetry breaking pattern, due to the singular behaviour
of the r.h.s.\ close to the origin: the crossing of right and left hand
sides of equation~(\ref{gap2}) does always occur and the value of $m$ where
this happens is a minimum  of the ef\/fective potential. In turn
the trivial solution, which at $H=0$ is a minimum, is transformed into
a maximum for $H\ne 0$.

Let us look at these dif\/ferent regimes more closely. As $H\to
0$, we recover the well-known NJL model gap equation
\begin{gather}
\label{stgap}
   1=\frac{GN_c}{2\pi^2}\left(J_0(m^2)
    +\frac{|QH|^2\Lambda^2}{3m^2(\Lambda^2+m^2)}+\cdots\right),
\end{gather}
where $J_0(m^2)$ is
\begin{gather*}
   J_0(m^2)=\Lambda^2 -m^2\ln\left(1+\frac{\Lambda^2}{m^2}\right).
\end{gather*}
Equation~(\ref{stgap}) at $H=0$ admits a nontrivial solution only if
$\tau >1$, where $\tau =G\Lambda^2N_c/2\pi^2$. This determines the
critical value $G_{\rm crit}=2\pi^2/\Lambda^2N_c$. The coupling $G$ is
said to be overcritical if $G>G_{\rm crit}$, and subcritical in the
opposite case when $G<G_{\rm crit}$.

At $m^2/\Lambda^2\ll 1$ the r.h.s.\ of equation (\ref{gap2}) is
\begin{gather}
         - \frac{|QH|}{\Lambda^2}\ln\left(
           \frac{m^2}{\Lambda^2}\right)+v(\xi )
           +{\cal O}\left(\frac{m^2}{\Lambda^2}\right).
\label{rhs}
\end{gather}
Here the function $v(\xi )$ of the argument $\xi =\Lambda^2/2|QH|$, is
given by
\begin{gather*}
   v(\xi )=\frac{1}{2\xi}\left[1-2\ln\Gamma\left(\xi
   +1\right)\right] + \psi\left(\xi\right).
\end{gather*}
This is a monotonically increasing function on the interval $0<\xi
<\infty$; $v(\xi )=0$ at the point $\xi\simeq 1.12$; the asymptotic
behaviour is
\begin{gather}
\label{asinf}
   v(\xi ) \sim 1-\frac{1}{2\xi}\ln (2\pi\xi),\qquad
                    \xi\to\infty , \\
   v(\xi ) \sim -\gamma -\frac{1}{2\xi}, \qquad \xi\to 0,
\label{as0}
\end{gather}
where $\gamma\simeq 0.577$ is the Euler's constant.

In the approximation considered the solution of equation~(\ref{gap2}) reads
\begin{gather*}
   M_{\rm dyn}=\Lambda \exp\left[-\xi\left(
            \frac{1}{\tau}-v(\xi )\right)\right].
\end{gather*}

To discuss the physical content of this result, we recall that the
energy spectrum of relativistic fermions in a constant magnetic
f\/ield $H$ contains Landau levels
\begin{gather*}
   E_n(p_z)=\pm\sqrt{\hat m^2+2|QH|n+p_z^2}, \qquad n=0,1,2,\ldots
\end{gather*}
with $p_z$ denoting the projection of the 3-momentum on the $z$-axis,
i.e., along the magnetic f\/ield. If the fermion mass $\hat
m$ goes to zero, as in the present case, there is no energy gap
between the vacuum and the LLL. Thus
the integer part of $\xi +1$ gives approximately the number of
Landau levels taken into account.

The f\/irst term in equation~(\ref{rhs}) has a clearly def\/ined
two-dimensional character with a logarithmic dependence on the
cutof\/f in the corresponding gap equation
\begin{gather}
\label{gap3}
   1=-\frac{GN_c}{2\pi^2}|QH|\ln\left(\frac{m^2}{\Lambda^2}\right)
\end{gather}
and, therefore, in the condensate (compare with equation~(\ref{stgap})).
Such  a behaviour is associated with the $(1+1)$-dimensional dynamics of
the fermion pairing on the energy surface $E_0=0$ of the LLL
\cite{Miransky:1995}. As long as this term dominates over the second
term, $v(\xi )$ in (\ref{rhs}), one concludes that the condensate is
mainly located on the LLL. Actually this condition is fulf\/illed
nearly everywhere at $\tau < 1$. This is obvious for $\xi =1$,
since $v(1) = 1/2+\psi (1) = 1/2 -\gamma\simeq -0.08$ is small compared
with $1/\tau$. For $\xi <1$ we come to the same conclusion after
considering the asymptotics of the second term (\ref{as0}). The other
formula, (\ref{asinf}), can be used to show that the above statement
is also true for $\xi >1$, except near the critical region $\tau\to
1-0$, where $v(\xi )$ dominates; then we must conclude that the
condensate spreads over many Landau levels.

In this special case it is possible to f\/ind an analytical
solution. Indeed, using (\ref{asinf}) in equation~(\ref{rhs}) we obtain
\begin{gather*}
   1-\frac{|QH|}{\Lambda^2}\ln\left(\frac{\pi m^2}{|QH|}\right)
   + {\cal O}\left(\frac{m^2}{\Lambda^2}, \frac{4|QH|^2}{\Lambda^4}
   \right).
\end{gather*}
We further suppose that the two following small
variables are of the same order
\begin{gather*}
  \frac{m^2}{\Lambda^2}\sim
  \left(\frac{|QH|}{\Lambda^2}\right)^2\sim\epsilon .
\end{gather*}
Then it follows immediately that the term with the logarithm is of
order $\sqrt{\epsilon}\ln\sqrt{\epsilon}$ and goes to zero, when
$\epsilon\to 0$. Thus, the gap equation
\begin{gather*}
   1-\frac{1}{\tau}= \frac{|QH|}{\Lambda^2}
   \ln\left(\frac{\pi m^2}{|QH|}\right) + {\cal O}(\epsilon )
\end{gather*}
is valid only in the region near the critical value $\tau\to 1-0$.
The closer $\tau$ to $1$, the smaller is $\epsilon$; Landau levels
approach a continuum distribution, and a condensate occupies many
levels. The physical reason for the changes found in the behaviour
of the condensate is the strength of the four-fermion interaction
which becomes essentially important here. The corresponding solution~is
\begin{gather*}
   M_{\rm dyn}^2=\frac{|QH|}{\pi}\exp\left[-\frac{\Lambda^2}{|QH|}
             \left(\frac{1}{\tau}-1\right)\right].
\end{gather*}

The near-critical regime which we f\/ind here dif\/fers from the result of
paper \cite{Miransky:1995}. In our case the dynamics of the fermion
pairing is essentially a $(3+1)$-dimensional one with a quadratic
dependence on the cutof\/f, to be compared with the previous case
in equation~(\ref{gap3}), where we had a $(1+1)$-dimensional behaviour with
only a logarithmic dependence on the cutof\/f.

\subsection[The extended multi-fermion system in a constant magnetic field]{The extended multi-fermion system in a constant magnetic f\/ield}\label{sec3.2}

Now we consider the three f\/lavour case with $\kappa,
g_1, g_2\neq 0$. In the simplest case with the octet f\/lavour
symmetry, where current quarks have equal masses
$\hat{m}_u=\hat{m}_d=\hat{m}_s$, which we set again zero, the system
(\ref{saddle-1}) reduces to a cubic equation with respect to
$h\equiv h_u=h_d=h_s$
\begin{gather}
\label{cubeq1}
   h^3 + \frac{\kappa}{12\lambda}\, h^2
         +\frac{4G}{3\lambda}\, h + \frac{4m}{3\lambda}=0
\end{gather}
with $\lambda =g_1+(2/3)g_2$. Making the replacement $h=\bar{h}-\kappa
/(36\lambda )$, one obtains from (\ref{cubeq1})
\begin{gather}
\label{cubeq2}
   \bar{h}^3 + t\bar{h} = b,
\end{gather}
where
\begin{gather*}
   t= \frac{4}{3} \left[ \frac{G}{\lambda} -
             \left(\frac{\kappa}{24\lambda}\right)^2\right], \qquad
   b= \frac{\kappa}{27\lambda}\left[
         \frac{G}{\lambda}-\frac{2}{3}\left(\frac{\kappa}{24\lambda}
         \right)^2\right] -\frac{4m}{3\lambda}.
\end{gather*}

\begin{figure}[t]
\centerline{\includegraphics[width=7.5cm]{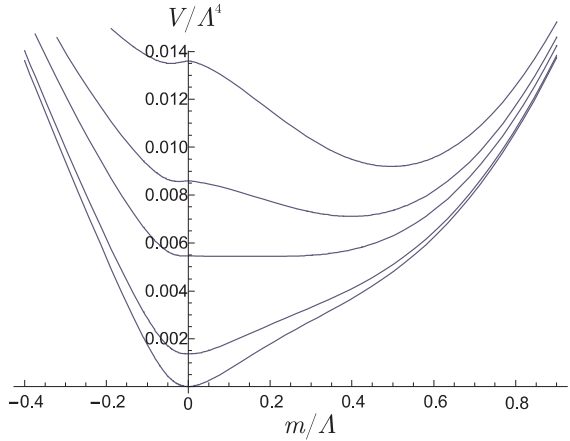}}
\vspace{-2mm}

\caption{The ef\/fective potential $V\Lambda^{-4}$ (in dimensionless units) for the parameters $G\Lambda^2=3$, $\kappa\Lambda^5=-800$, $\lambda \Lambda^8= 1.667\cdot 10^3$. From bottom to top the curves correspond to $|QH|\Lambda^{-2}=0.05$; $0.1;\, 0.2275;\, 0.3;\, 0.4$. }
\label{fig2}\vspace{-2mm}
\end{figure}

This cubic equation has one real root, if $t>0$, i.e.,
\begin{gather}
\label{stabcond}
   \frac{G}{\lambda}>\left(\frac{\kappa}{24\lambda}\right)^2,
\end{gather}
which is a particular case of the inequalities (\ref{ineq1}).
Assuming that the couplings fulf\/ill condition~(\ref{stabcond}),
we f\/ind (at any given value of $b$) the real solution of equation
(\ref{cubeq2})
\begin{gather*}
   \bar h (m)= \left(\frac{b}{2}+\sqrt{D}\right)^{\frac{1}{3}}
        + \left(\frac{b}{2}-\sqrt{D}\right)^{\frac{1}{3}}  ,
   \qquad
   D= \left(\frac{t}{3}\right)^3+\left(\frac{b}{2}\right)^2,
\end{gather*}
with the important property $\bar h(0)=\kappa /36\lambda$ which
ensures that the trivial solution of the gap equation, $m=0$, exists
(see~\cite{Osipov:2007} for more details about this part of the
potential).

In Fig.~\ref{fig2}  the ef\/fective potential is shown for the parameters $G\Lambda^2=3$, $\kappa\Lambda^5=-800$, $\lambda \Lambda^8= 1.67\cdot 10^3$. The catalyzed minima close to the origin at small values of the magnetic f\/ield get washed out with increasing strength of $H$ giving way to a second minimum at larger values of~$M_{\rm dyn}$. For the lower values of $H$, the catalyzed minimum is so close to the origin that one can not discern it from the trivial maximum which always occurs at the origin (see also discussion around equation~(\ref{sder})). The catalysis of dynamical symmetry breaking can again most easily be seen by considering the non trivial solutions of the corresponding gap equation, to which we turn now.

\begin{figure}[t]
\centerline{\includegraphics[width=6.91cm]{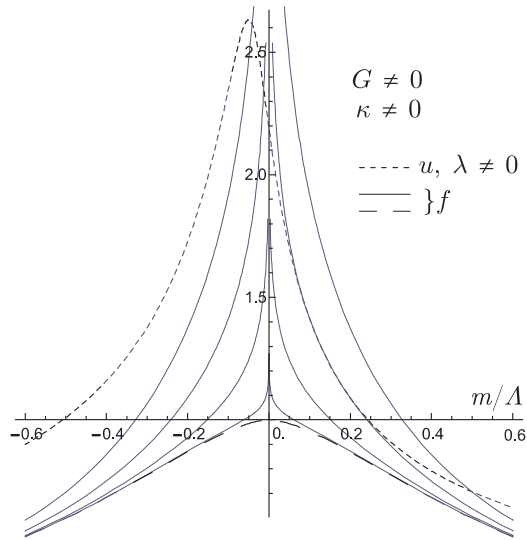}}
\vspace{-2mm}

\caption{The l.h.s.\ (short dashed line, label $u$) and r.h.s.\ (label $f$) of the gap equation (all in dimensionless units) with the parameter set $G$, $\kappa$, $\lambda$ of Fig.~\ref{fig2}. From bottom to top the r.h.s.\ curves have $|QH|\Lambda^{-2}=0$ (long dashed curve),
$|QH|\Lambda^{-2}=0.02;\, 0.1;\, 0.2275;\ 0.4$ (solid curves). The intersecting solutions $M_{\rm dyn}$ are displayed in Fig.~\ref{fig4}.}
\label{fig3}\vspace{-2mm}
\end{figure}

\begin{figure}[t]
\centerline{\includegraphics[width=7.9cm]{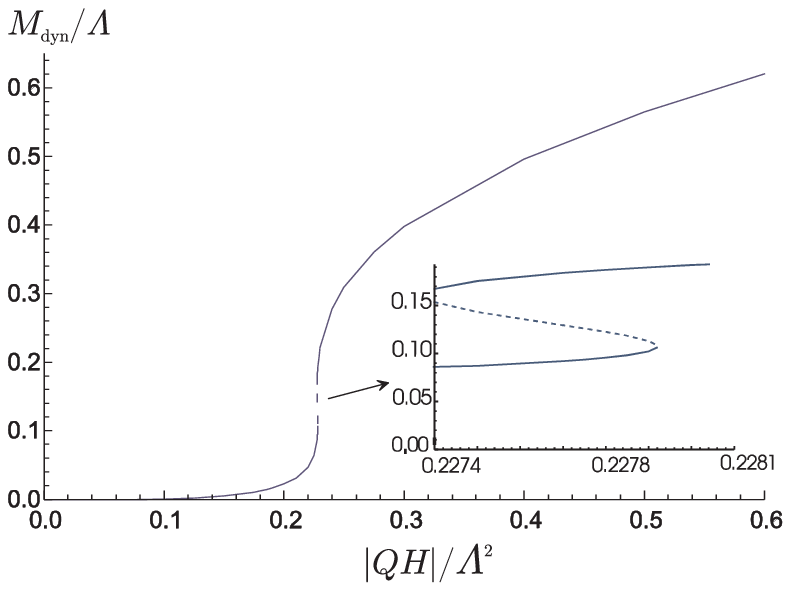}}
\vspace{-2mm}

\caption{The non-trivial solutions  $M_{\rm dyn}/\Lambda$ of the gap equation, as functions of $|QH|\Lambda^{-2}$, for the parameter set $G$, $\kappa$, $\lambda$ of Figs.~\ref{fig2} and~\ref{fig3}. For $0<|QH|\Lambda^{-2}< 0.2274$, $M_{\rm dyn}$  corresponds to the catalysis phase. Then a rapid transition leads the system to the larger $M_{\rm dyn}$ solution induced by the higher order multi-quark forces. A coexistence of both kinds of solutions (at the interrupted line) happens in a very narrow window of $|QH|\Lambda^{-2}$ values (see box insertion: short dashed line are maxima, solid lines minima). All quantities are dimensionless.}
\label{fig4}\vspace{-2mm}
\end{figure}

In a constant magnetic f\/ield the gap equation is
determined by
\begin{gather}
\label{gap5}
  -\frac{2\pi^2h(m)}{\Lambda^2N_cm}= f(m^2;\Lambda,|QH|).
\end{gather}
Comparing this result with equation~(\ref{gap2}), one sees that only the
l.h.s. is changed. The six- and eight-quark interactions have
modif\/ied it in such a way that now we get a function $h(m)/m$
instead of the former constant term involving only the coupling of
four-quark interactions, $-1/G$. The l.h.s.\ of (\ref{gap5}),
abbreviated by $u$ in Fig.~\ref{fig3}, has now a bell-shaped form
(short-dashed line), to be compared with the horizontal line of
Fig.~\ref{fig1}. Since $h(m)/m =-1/G+{\cal O}(m)$, the
bell-shaped curve crosses the ordinate axis at the same point as the
former straight line (for the same value of $G\Lambda^2$). The r.h.s.\
is again represented by the long-dashed curve ($H=0$ case) and by the
full lines for the f\/inite $H$ cases. As mentioned before these
are not altered by the couplings~$G$,~$\kappa$,~$\lambda$. The
intersection points of the l.h.s.\ with the r.h.s.\ curves yield the
nontrivial solutions of the gap equation: either one or
three solutions can be found for $m>0$. For the parameter set considered the
region of coexisting solutions is very narrow, see Fig.~\ref{fig4}.
If equation~(\ref{gap5}) has no non-trivial
solutions at $H=0$ (the present case) we say that the set of couplings $G$, $\kappa$, $\lambda$ is subcritical. It is said to be overcritical in the
opposite case. Note that the overcritical set may contain $G<G_{\rm crit}$ (a parameter set for this case can be found in~\cite{Osipov:2007b}; there the coexistence of solutions f\/ills a larger interval in $H$, and starts at $H=0$).

The trivial solution, $m=0$, corresponds to the point where the
potential, $V(m)$, reaches its local maximum. The second
derivative
\begin{gather}
\label{sder}
   \lim_{m\to 0} \frac{d^2V(m)}{dm^2}=\lim_{m\to 0}
   \frac{N_c|QH|}{2\pi^2}\ln\frac{|m|\Lambda}{2|QH|}=-\infty
\end{gather}
is negative here. This is the general mathematical reason for the
phenomena known as magnetic catalysis of dynamical f\/lavour
symmetry breaking. The logarithmically divergent negative result
ensures that this phase transition always takes place, if $H\neq 0$.
This transition does not depend on the details related with the multi-quark
dynamics,  i.e., the result is true even
for free fermions in a constant magnetic f\/ield.

\looseness=-1
What is really dependent on the multi-quark dynamics is the local
minima structure of the theory. Let us recall that in the theory
with just four-fermion interactions the ef\/fective potential has
only one minimum at $m>0$, and this property does not depend on the
strength of the f\/ield~$H$. We have demonstrated this in
Fig.~\ref{fig1}. In the theory with four-, six-, and eight-quark
interactions one can f\/ind either one or two local minima at
$m>0$. The result depends on the strength of the magnetic f\/ield
$H$, and couplings $G$, $\kappa$, $\lambda$. This is illustrated in
Figs.~\ref{fig2}--\ref{fig4}.
Namely, the upper full curve (r.h.s.\ of equation~(\ref{gap5}) for $|QH|\Lambda^{-2}=0.4$) has only one intersection
point with the bell-shaped curve $u$ (l.h.s.\ of equation~(\ref{gap5}) for
$G\Lambda^2=3$, $\kappa\Lambda^5=-800$, $\lambda\Lambda^8=1667.$), as seen in Fig.~\ref{fig3}. This
point corresponds to a single vacuum state of the theory at a large $M_{\rm dyn}$. As opposed to this the curve with
$|QH|\Lambda^{-2}=0.2275$
has three intersections with the same curve~$u$; these intersections,
taken in order, correspond to a local minimum, a local maximum and
a further local minimum of the potential (see also Fig.~\ref{fig4}). Reducing further the f\/ield strength one obtains again only one solution, at considerable smaller values of $M_{\rm dyn}$, corresponding to catalysis.

Therefore the f\/irst minimum catalyzed by
a constant magnetic f\/ield is then smoothed out with increasing $H$. It ceases to
exist at some critical value of $|QH|\Lambda^{-2}$, from which on only
the large $M_{\rm dyn}$ solution survives,  as shown in Fig.~\ref{fig4},
for the parameter set of Figs.~\ref{fig2},~\ref{fig3}. This process is accompanied by
a sharp increase in depth of the ef\/fective potential at the
second minimum (see Fig.~\ref{fig2}).
It is clear that the only way to wash out the
f\/irst minimum (due to  equation~(\ref{sder})), is by lowering the
barrier between this state and the second minimum.
We also observe that the second
minimum is unremovable, because the asymptotic behaviour of the
functions in equation~(\ref{gap5}) is such that the l.h.s. dominates over
the r.h.s.\ at large $m/\Lambda$. Let us stress that this
scenario will always be possible if at $H=0$ the six- and eight-quark
interactions induce dynamical symmetry breaking in the subcritical
regime $G<G_{\rm crit}$ of four-quark interactions (the case presented in~\cite{Osipov:2007b}).
If the system remains subcritical also for the higher-order interactions, as in the example
shown here, the regime of coexistence of more than one minimum of the gap equation at f\/inite $H$ shrinks eventually to zero by lowering further the eight-quark interaction strength $\lambda$ (keeping the remaining parameters f\/ixed). Then one can still observe a crossover transition from small to large $M_{\rm dyn}$ solutions. For example by changing the r.h.s.\ of the stability condition $G-\left(\frac{\kappa}{24}\right)^2\frac{1}{\lambda} >0$ from $2.33$ (parameter set shown in Figs.~\ref{fig2}--\ref{fig4}) to~$1.$, one obtains qualitatively the same picture as in Fig.~\ref{fig4} (of course without coexistence region) but a smoother transition.

\section{Concluding remarks}\label{sec4}

The ef\/fects of interactions among fermions on catalysis of dynamical
chiral symmetry breaking by a constant magnetic f\/ield have been considered.
A generalized NJL model with $U(3)$ f\/lavor symmetry containing up to
eight-quark interactions has been at the basis of this study, and the motivation
for the use of such a Lagrangian was brief\/ly reviewed.  A critical
survey of the non-trivial solutions of the gap equations has been obtained
and compared to the known solutions of the conventional four-quark interaction NJL model.

We have discussed two cases: (i) all coupling strengths are
subcritical, (ii) the addition of six and eight quark interactions
transform the original subcritical $G<G_{\rm crit}$ four-quark system into an overcritical one, inducing the formation of a large condensate.

In both cases we have that in the region $m^2/\Lambda^2\ll 1$ the four-fermion interaction
dominates the behaviour of the system. Since their coupling strength
is small, $G<G_{\rm crit}$, the massless fermions behave like almost free
particles moving in a weak external magnetic f\/ield, with access
to a large number of Landau levels, $\xi\gg 1$. This f\/ield
catalyzes the process of fermion-antifermion pairing on the energy
surface $E_0=0$ of the LLL. The f\/irst minimum localized at
$m^2/\Lambda^2\ll 1$ is exactly formed by such an $(1+1)$-dimensional
condensate. If six- and eight-fermion forces would not act on the
system, this ground state would be stable: our formulae as well as the
result of paper \cite{Miransky:1995} show clearly that a slow increase
of the strength $H$ does not wash out the condensate from the energy
surface $E_0=0$ of the LLL. However, when the six- and eight-fermion
interactions are present, a slowly increasing magnetic f\/ield
destroys f\/inally this ground state. In case (ii) this always happens;
in case (i) the transition can either be sharp as in (ii) or of a crossover type,
depending on the interaction strengths. The new condensate has a~$(3+1)$-dimensional structure similar in every respect to the standard
NJL case with broken chiral symmetry at $H=0$,   i.e., when the
condensate spreads over many single fermion states. This is because
the increasing magnetic f\/ield enlarges the dynamical fermion
mass, and scales of order $m/\Lambda\sim 1$ become relevant. At these
scales the 't Hooft and eight-quark interactions push the system to a
new regime, where the fermions are not anymore free-like
particles: they  interact strongly with each other and this
interaction changes essentially the fermionic spectrum and the
structure of the ground state with all the above mentioned
consequences.

Thus we have obtained not only a correct description of the
well-known physics related with the LLL, but have found also a clear
signature for the possibly important role played by 't Hooft and
eight-quark interactions. Namely, in the presence of these
interactions the magnetic f\/ield can change the condensation
zone from the zero-energy surface of the LLL to a wide region spread
over many Landau levels and vice versa. One can expect that hard gamma
emissions accompany this process.

The value of the characteristic scale $\Lambda$ has not been specif\/ied yet.
We assume that this value is determined by the problem under consideration. Its
choice can also be motivated by the number of Landau levels to be
considered.
The characteristic scale of the magnetic f\/ields which
can induce such a transition is  of the order $H=7\cdot 10^{13}\,
\Lambda^2\,\mbox{Gauss/MeV$^2$}$s, and actually depend on the
cutof\/f involved in the problem. For instance, in hadronic
matter it is probably reasonable to assume that $\Lambda\simeq 800\,
\mbox{MeV}$, leading to $H=4.5\cdot 10^{19}\,\mbox{G}$.

A potentially interesting area where this
ef\/fect may f\/ind applications are studies of compact stellar objects in presence of strong
magnetic f\/ields, in particular the young neutron stars,
magnetars~\cite{Duncan:1992}. The surface magnetic f\/ields are
observed to be $\geq 10^{15}$~G, but actually they can be even
much higher at the core region. Another area is connected with the
electroweak phase transition in the early Universe \cite{Olesen:1992},
where the strength of magnetic f\/ields can reach
$H\sim 10^{24}$~G.
Even giving conservative limitations on the primordial magnetic f\/ield of present epoch value of~$10^{-9}$~G, based on primordial nuclear synthesis calculations \cite{BB}, it seems clear that the magnetic f\/ield reached the transition regime discussed in this paper in the past.  A~very rough estimate based on the relation $H\propto T_\nu^2$ ($T_\nu$ being the neutrino temperature), gives a~magnetic f\/ield strength at baryogenesis of at least $10^{17}$~G, of $10^{21}$~G at the epoch of electroweak breaking and of $10^{45}$~G at the GUT scale. The coupling to those strong magnetic f\/ields will have inf\/luence on those processes. However to incorporate the f\/indings of the present work into those mechanisms is beyond the scope of the present article.

\subsection*{Acknowledgements}
B. Hiller is very grateful to the organizers for the kind invitation and hospitality
and for creating the conditions for a very interesting conference.
This work has been supported in part by grants provided by
Funda\c c\~ao para a Ci\^encia e a Tecnologia, POCI 2010 and FEDER,
POCI/FP/63930/2005, POCI/FP/81926/2007.  This research is part of the EU integrated
infrastructure initiative Hadron Physics project under contract
No.RII3-CT-2004-506078.

\pdfbookmark[1]{References}{ref}
\LastPageEnding
\end{document}